\documentclass{ifacconf}

\usepackage{graphicx}      
\usepackage{natbib}        

\usepackage[latin1]{inputenc}
\usepackage[usenames,dvipsnames]{pstricks}
\usepackage{amsmath}
\usepackage{amssymb}

\begin{document}
\begin{frontmatter}

\title{\mbox{Grey-box nonlinear state-space modelling} for mechanical vibrations identification} 

\author[First,Second]{J.P. No\"el} 
\author[First]{J. Schoukens} 
\author[Second]{G. Kerschen}

\address[First]{Department ELEC \\
Vrije Universiteit Brussel, Brussels, Belgium \\
e-mail: jp.noel@ulg.ac.be, johan.schoukens@vub.ac.be}
\vspace{0.25cm}
\address[Second]{Space Systems and Structures Laboratory (S3L) \\
Aerospace and Mechanical Engineering Department \\
University of Li\`ege, Li\`ege, Belgium \\
e-mail: g.kerschen@ulg.ac.be}

\begin{abstract}                
In the present paper, a flexible and parsimonious model of the vibrations of nonlinear mechanical systems is introduced in the form of state-space equations. It is shown that the nonlinear model terms can be formed using a limited number of output measurements. A two-step identification procedure is derived for this grey-box model, integrating nonlinear subspace initialisation and maximum likelihood optimisation. The complete procedure is demonstrated on the Silverbox benchmark, which is an electrical mimicry of a single-degree-of-freedom mechanical system with one displacement-dependent nonlinearity.
\end{abstract}


\end{frontmatter}

\section{Introduction}

Nonlinear system identification constantly faces the compromise between the flexibility of the fitted model and its parsimony. Flexibility refers to the ability of the model to capture complex nonlinearities, while parsimony is its quality to possess a low number of parameters. In this regard, a nonlinear state-space representation
\begin{equation}
\left\lbrace
\begin{array}{r c l}
    \mathbf{\dot{x}}(t) & = & \mathbf{A} \: \mathbf{x}(t) + \mathbf{B} \: \mathbf{u}(t) + \mathbf{E} \: \mathbf{g}(\mathbf{x},\mathbf{u}) \\
    \mathbf{y}(t) & = & \mathbf{C} \: \mathbf{x}(t) + \mathbf{D} \: \mathbf{u}(t) + \mathbf{F} \: \mathbf{h}(\mathbf{x},\mathbf{u})
\end{array} \right.
\label{Eq:BlackBoxStateSpace}
\end{equation}
can be classified as very flexible but little parsimonious, two features typically shared by black-box models. In Eqs.~(\ref{Eq:BlackBoxStateSpace}), $\mathbf{A} \in \mathbb{R}^{\: n_{s} \times n_{s}}$, $\mathbf{B} \in \mathbb{R}^{\: n_{s} \times m}$, $\mathbf{C} \in \mathbb{R}^{\: l \times n_{s}}$ and $\mathbf{D} \in \mathbb{R}^{\: l \times m}$ are the linear state, input, output and direct feedthrough matrices, respectively; $\mathbf{x}(t) \in \mathbb{R}^{\: n_{s}}$ is the state vector; $\mathbf{y}(t) \in \mathbb{R}^{\: l}$ and $\mathbf{u}(t) \in \mathbb{R}^{\: m}$ are the output and input vectors, respectively. The linear-in-the-parameters expressions $\mathbf{E} \: \mathbf{g}(\mathbf{x},\mathbf{u}) \in \mathbb{R}^{\: n_{s}}$ and $\mathbf{F} \: \mathbf{h}(\mathbf{x},\mathbf{u}) \in \mathbb{R}^{\: l}$ are the nonlinear model terms coupling the state and input variables. The order of the model, \textit{i.e.} the dimension of the state space, is noted $n_{s}$.

In the present paper, it is shown that, in the case of mechanical systems where nonlinearities are physically localised, the model structure in Eqs.~(\ref{Eq:BlackBoxStateSpace}) can be drastically simplified. More specifically, Section 2 demonstrates that the nonlinear terms in Eqs.~(\ref{Eq:BlackBoxStateSpace}) can be constructed using a limited number of output measurements, and so without resorting to the state and input vectors. This makes the resulting grey-box state-space model a parsimonious representation of nonlinear mechanical systems. An efficient, two-step identification procedure for this model is derived in Section 3, integrating nonlinear subspace initialisation and maximum likelihood optimisation. Finally, the complete procedure is applied in Section 4 to an electrical circuit mimicking the behaviour of a single-degree-of-freedom mechanical system with one displacement-dependent nonlinearity.

\section{Grey-box nonlinear state-space modelling of mechanical vibrations}

In the analysis of mechanical vibrations, one very often distinguishes localised nonlinearities, which are physically confined to a small area, from nonlinearities distributed throughout (some large region of) the entire structure. Localised elements are arguably the most common in engineering practice, since structural nonlinearities typically arise from the complex dynamics of joints interfacing subcomponents. Many meaningful examples of this reality are to be found in the aerospace sector. For instance, nonlinearities resulting from the appearance of gaps in the truss supports of the Huygens probe were attested during the modal survey of the Cassini spacecraft~[\cite{Carney_IMAC1997}]. Nonlinearities were also reported during ground vibration testing of the Airbus A400M, and were attributed to the elastomeric mounts supporting the four turboprop engines of the aircraft~[\cite{Ahlquist_A400M_IMAC2010}]. Moreover, the analysis of in-orbit data of the International Space Station highlighted that the opening of a pin connection in the assembly of its solar arrays led to severe nonlinearity~[\cite{Laible_IMAC2013}].

Assuming localised nonlinearities, the vibrations of a $n_{p}$-degree-of-freedom mechanical system obey Newton's second law written in the form
\begin{equation}
\mathbf{M} \: \mathbf{\ddot{q}}(t) + \mathbf{C}_{v} \: \mathbf{\dot{q}}(t) + \mathbf{K} \: \mathbf{q}(t) + \displaystyle \sum^{s}_{a=1} {c_{a} \: \mathbf{g}_{a}(\mathbf{q}_{nl}(t),\mathbf{\dot{q}}_{nl}(t))} = \mathbf{p}(t) ,
\label{Eq:Newton}
\end{equation}
where $\mathbf{M}$, $\mathbf{C}_{v}$, $\mathbf{K} \in \mathbb{R}^{\: n_{p} \times n_{p}}$ are the mass, linear viscous damping and linear stiffness matrices, respectively; $\mathbf{q}(t)$ and $\mathbf{p}(t) \in \mathbb{R}^{\: n_{p}}$ are the generalised displacement and external force vectors, respectively; the nonlinear restoring force term is written using $s$ basis function vectors $\mathbf{g}_{a}(t) \in \mathbb{R}^{\: n_{p}}$ associated with coefficients $c_{a}$. The subset of generalised displacements and velocities involved in the construction of the basis functions are denoted $\mathbf{q}_{nl}(t)$ and $\mathbf{\dot{q}}_{nl}(t)$, respectively. Physically, they correspond to mechanical degrees of freedom located on both sides of the localised nonlinearities in the system.

\begin{figure}[ht]
\begin{center}
    \begin{pspicture}(9,5)
    \psframe[linewidth=0.5pt,dimen=outer](2.8,0.5)(6.4,2) 
    \psframe[linewidth=0.5pt,dimen=outer](2.8,3.0)(6.4,4.5) 
		\psline[arrows=->,linewidth=0.5pt,arrowinset=0,arrowlength=1.2,arrowsize=3pt 2](0,3.75)(1.2,3.75) 
		\psline[arrows=->,linewidth=0.5pt,arrowinset=0,arrowlength=1.2,arrowsize=3pt 2](1.8,3.75)(2.8,3.75) 
		\psline[arrows=->,linewidth=0.5pt,arrowinset=0,arrowlength=1.2,arrowsize=3pt 2](6.4,3.75)(8.8,3.75) 
		\psline[arrows=->,linewidth=0.5pt,arrowinset=0,arrowlength=1.2,arrowsize=3pt 2](7.7,1.25)(6.4,1.25) 
		\psline[arrows=->,linewidth=0.5pt,arrowinset=0,arrowlength=1.2,arrowsize=3pt 2](1.5,1.25)(1.5,3.45) 
		\psline[linewidth=0.5pt](7.7,3.75)(7.7,1.25) 
    \psline[linewidth=0.5pt](2.8,1.25)(1.5,1.25) 
    \rput[bl](3.25,3.8){Underlying linear}
    \rput[bl](3.15,3.2){system: $\mathbf{M}$, $\mathbf{C}_{v}$, $\mathbf{K}$}
    \rput[bl](3.05,1.4){Nonlinear feedback:}
    \rput[bl](3.05,0.7){$c_{a}$, $\mathbf{g}_{a}(\mathbf{q}_{nl}(t),\mathbf{\dot{q}}_{nl}(t))$}
		\rput[bl](0.2,3.9){$\mathbf{p}(t)$}
    \rput[bl](7.0,3.9){$\mathbf{q}(t)$, $\mathbf{\dot{q}}(t)$}
		\pscircle[linewidth=0.5pt,dimen=outer](1.5,3.75){0.3}
    \rput[bl](1.34,3.62){+}
    \end{pspicture}
\caption{Feedback interpretation of Newton's law in Eq.~(\ref{Eq:Newton}).}
\label{Fig:Feedback}
\end{center}
\end{figure}
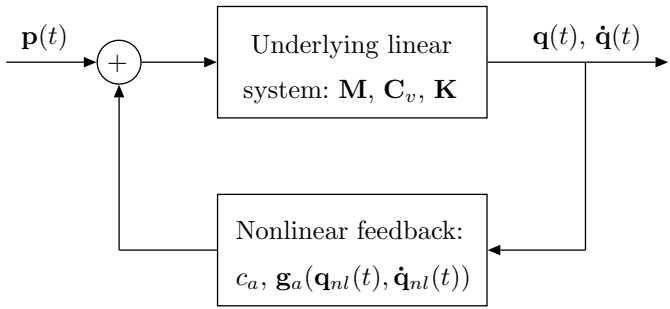

The dynamics governed by Eq.~(\ref{Eq:Newton}) is conveniently interpreted by moving the nonlinear restoring force term to the right-hand side, \textit{i.e.}
\begin{equation}
\mathbf{M} \: \mathbf{\ddot{q}}(t) + \mathbf{C}_{v} \: \mathbf{\dot{q}}(t) + \mathbf{K} \: \mathbf{q}(t) = \mathbf{p}(t)  - \displaystyle \sum^{s}_{a=1} {c_{a} \: \mathbf{g}_{a}(\mathbf{q}_{nl}(t),\mathbf{\dot{q}}_{nl}(t))} ,
\label{Eq:Newton2}
\end{equation}
which leads to the block diagram in Fig.~\ref{Fig:Feedback}.

The feedback structure of this diagram suggests that localised nonlinearities in mechanical systems act as additional inputs applied to the underlying linear system. This, in turn, reveals that black-box nonlinear terms in a state-space model, such as $\mathbf{E} \; \mathbf{g}(\mathbf{x},\mathbf{u})$  and $\mathbf{F} \; \mathbf{h}(\mathbf{x},\mathbf{u})$ in Eqs.~(\ref{Eq:BlackBoxStateSpace}), are overly complex to address mechanical vibrations. A more parsimonious description of nonlinearities is achieved by translating Eq.~(\ref{Eq:Newton2}) in state space, which provides the grey-box model
\begin{equation}
\left\lbrace
\begin{array}{r c l}
    \mathbf{\dot{x}}(t) & = & \mathbf{A} \: \mathbf{x}(t) + \mathbf{B} \: \mathbf{u}(t) + \mathbf{E} \: \mathbf{g}(\mathbf{y}_{nl}(t),\mathbf{\dot{y}}_{nl}(t)) \\
    \mathbf{y}(t) & = & \mathbf{C} \: \mathbf{x}(t) + \mathbf{D} \: \mathbf{u}(t) + \mathbf{F} \: \mathbf{g}(\mathbf{y}_{nl}(t),\mathbf{\dot{y}}_{nl}(t)) , \\
\end{array} \right.
\label{Eq:GreyBoxStateSpace0}
\end{equation}
where $\mathbf{g}(t) \in \mathbb{R}^{\: s}$ is a vector concatenating the nonzero elements in the basis function vectors $\mathbf{g}_{a}(t)$, and $\mathbf{E} \in \mathbb{R}^{\: n_{p} \times s}$ and $\mathbf{F} \in \mathbb{R}^{\: l \times s}$ are the associated coefficient matrices; $\mathbf{y}_{nl}(t)$ and $\mathbf{\dot{y}}_{nl}(t)$ are the subsets of the measured displacements and velocities located close to nonlinearities, respectively.

For the sake of conciseness, one adopts the concatenated equations
\begin{equation}
\left\lbrace
\begin{array}{r c l}
    \mathbf{\dot{x}}(t) & = & \mathbf{A} \: \mathbf{x}(t) + \overline{\mathbf{B}} \: \overline{\mathbf{u}}(t) \\
    \mathbf{y}(t) & = & \mathbf{C} \: \mathbf{x}(t) + \overline{\mathbf{D}} \: \overline{\mathbf{u}}(t) , \\
\end{array} \right.
\label{Eq:GreyBoxStateSpace}
\end{equation}
where $\overline{\mathbf{B}} = \left[\mathbf{B} \ \ \mathbf{E}\right]$ and $\overline{\mathbf{D}} = \left[\mathbf{D} \ \ \mathbf{F}\right]$; the extended input vector $\overline{\mathbf{u}}(t)$ is similarly defined as $\left[\mathbf{u}(t)^{T} \ \ \mathbf{g}(t)^{T}\right]^{T}$, where $T$ is the transpose operation.

\section{Identification procedure}

The simplified, linear-like structure of the grey-box state-space model given in Eqs.~(\ref{Eq:GreyBoxStateSpace}) lends itself to an efficient, two-step identification procedure. First, initial estimates of the $\mathbf{A}$, $\overline{\mathbf{B}}$, $\mathbf{C}$ and $\overline{\mathbf{D}}$ matrices are obtained using a nonlinear subspace identification method. Second, the subspace estimates are optimised in maximum likelihood sense by applying a nonlinear minimisation routine. The complete procedure is carried out in the frequency domain, which opens the possibility to apply user-defined weighting functions in specific frequency bands. For the sake of comparison, the identification of the black-box polynomial nonlinear state-space model in Ref.~[\cite{Paduart_PNLSS}] requires a four-step process, including linear subspace parameter estimation and two nonlinear optimisation searches. 

\subsection{Initialisation using a nonlinear subspace method}

Subspace algorithms are well known for solving linear system identification problems~[\cite{VODM_Book,McKelvey_Subspace}]. In recent years, nonlinear generalisations of subspace identification have also emerged, following the original idea of Lacy and Bernstein~[\cite{Lacy_Subspace}]. The present study exploits the frequency-domain nonlinear subspace method proposed in Ref.~[\cite{JPNoel_FNSI}], termed FNSI method, to calculate an initial set of matrices $\left(\mathbf{A},\overline{\mathbf{B}},\mathbf{C},\overline{\mathbf{D}} \right)$. A consistent estimate of the set is obtained if both the input $\mathbf{u}(t)$ and the nonlinear basis functions in $\mathbf{g}(\mathbf{y}_{nl}(t),\mathbf{\dot{y}}_{nl}(t))$ are noiseless, as inferred from the results in Refs.~[\cite{McKelvey_Subspace,Pintelon_Subspace}]. The use of FNSI offers two important advantages. Firstly, the method constructs a fully nonlinear model of the system under test from the beginning, in contrast with the linearised model which serves as a starting point in Ref.~[\cite{Paduart_PNLSS}]. Secondly, the method embeds an intuitive model order selection capability via stabilisation diagrams~[\cite{JPNoel_TFNSI}]. 

\subsection{Nonlinear optimisation of the subspace model in the maximum likelihood framework}

In the case of output measurements with a low signal-to-noise ratio, consistency of the initial state-space matrices $\left(\mathbf{A},\overline{\mathbf{B}},\mathbf{C},\overline{\mathbf{D}} \right)$ is lost. Unbiased parameter estimates can be recovered by optimising the subspace model in maximum likelihood (ML) sense. The ML framework also guarantees the lowest possible uncertainty on the model parameters, \textit{i.e.} the efficiency of the estimates~[\cite{JSchoukens_Book}]. To formulate the ML cost function, the state-space Eqs.~(\ref{Eq:GreyBoxStateSpace}) are recast in the frequency domain as
\begin{equation}
\left\lbrace
\begin{array}{r c l}
    \xi_k \: \mathbf{X}(k) & = & \mathbf{A} \: \mathbf{X}(k) + \overline{\mathbf{B}} \: \overline{\mathbf{U}}(k) \\
    \mathbf{Y}(k) & = & \mathbf{C} \: \mathbf{X}(k) + \overline{\mathbf{D}} \: \overline{\mathbf{U}}(k) ,
\end{array} \right.
\label{Eq:FDStateSpace}
\end{equation}
where $k$ is the frequency line, $\xi_k$ the Laplace and z-transform variable depending on whether a continuous- or discrete-time formulation is selected, and $\mathbf{Y}(k)$, $\mathbf{X}(k)$ and $\overline{\mathbf{U}}(k)$ the discrete Fourier transforms (DFTs) of $\mathbf{y}(t)$, $\mathbf{x}(t)$ and $\overline{\mathbf{u}}(t)$, respectively.

\vspace*{0.25cm}
\textbf{Assumption on the noise model}

The extended input spectrum is assumed to be noiseless, \textit{i.e.} observed without errors and independent of the output noise. The output disturbing noise term $\mathbf{N}_{Y}(k)$ is Gaussian distributed, has zero mean $\mathcal{E} \left (\mathbf{N}_{Y}(k) \right) = 0$, where $\mathcal{E}$ is the expectation operator, and has a covariance matrix with only nonzero diagonal elements equal to $\boldsymbol{\sigma}^{2}_{Y}(k) = \mathcal{E} \left ( \left| \mathbf{N}_{Y}(k) \right|^{2} \right)$, as described in Ref.~[\cite{JSchoukens_Noise}].

\vspace*{0.25cm}
\textbf{Maximum likelihood cost function}

The Gaussianity of the output noise implies that the ML cost function simplifies to a weighted least-squares estimator in the frequency domain~[\cite{JSchoukens_Book}]. Introducing the vector of model parameters $\theta$ as
\begin{equation}
\theta = \left[vec\left(\mathbf{A}\right) \:;\: vec\left(\overline{\mathbf{B}}\right) \:;\: vec\left(\mathbf{C}\right) \:;\: vec\left(\overline{\mathbf{D}}\right) \right] ,
\label{Eq:theta}
\end{equation}
where the operation denoted $vec$ stacks the columns of a matrix on top of each other, the cost function to minimise hence writes
\begin{equation}
\mathbf{V}(\theta) = \displaystyle \sum^{F}_{k=1} \boldsymbol{\epsilon}^{H}(k,\theta) \: \mathbf{W}^{2}(k) \: \boldsymbol{\epsilon}(k,\theta) ,
\label{Eq:WLS}
\end{equation}
where $F$ is the number of processed lines, $H$ the Hermitian transpose, and $\mathbf{W}(k)$ a weighting function chosen equal to $\boldsymbol{\sigma}^{-1}_{Y}(k)$. The model error vector $\boldsymbol{\epsilon} \in \mathbb{R}^{\: l}$ is defined as the complex-valued difference
\begin{equation}
\boldsymbol{\epsilon}(k,\theta) = \mathbf{Y}_{m}(k,\theta) - \mathbf{Y}(k) ,
\label{Eq:epsilon}
\end{equation}
where $\mathbf{Y}_{m}(k,\theta)$ and $\mathbf{Y}(k)$ are the DFTs of the modelled and measured outputs, respectively.

\vspace*{0.25cm}
\textbf{Analytical calculation of the Jacobian matrix}

In practice, the nonlinear least-squares cost function in Eq.~(\ref{Eq:WLS}) is minimised using a Levenberg-Marquardt optimisation algorithm, which combines the large convergence region of the gradient descent method with the fast convergence of the Gauss-Newton method~[\cite{Levenberg,Marquardt}]. This algorithm requires the calculation of the Jacobian matrix $\mathbf{J}(k,\theta)$ associated with the cost function or, equivalently, with the error function in Eq.~(\ref{Eq:epsilon}), \textit{i.e.}
\begin{equation}
\mathbf{J}(k,\theta) = \frac{\partial \boldsymbol{\epsilon}(k,\theta)}{\partial \theta} = \frac{\partial \mathbf{Y}_{m}(k,\theta)}{\partial \theta} .
\label{Eq:JWLS}
\end{equation}
Given the nonlinear relationship which exists between $\mathbf{Y}(k)$ and $\overline{\mathbf{U}}(k)$, it may not be practical to compute the elements of $\mathbf{J}(k,\theta)$ directly in the frequency domain. An alternative approach consists in carrying out the computation of the Jacobian matrix in the time domain, and then in applying the DFT. One first focuses on the determination of the element $J_{A_{ij}}(t) \in \mathbb{R}^{\: l}$ of the time-domain Jacobian defined as
\begin{equation}
J_{A_{ij}}(t) = \dfrac{\partial \mathbf{y}(t)}{\partial A_{i \: j}} .
\label{Eq:JAij_Def}
\end{equation}

The derivative of the output relation in Eqs.~(\ref{Eq:GreyBoxStateSpace}) with respect to $A_{i \: j}$ is given by
\begin{equation}
\begin{array}{r c l}
    \dfrac{\partial \mathbf{y}(t)}{\partial A_{i \: j}} & = & \dfrac{\partial}{\partial A_{i \: j}} \left ( \mathbf{C} \: \mathbf{x}(t) + \overline{\mathbf{D}} \: \overline{\mathbf{u}}(t) \right) \\
     & = & \mathbf{C} \: \dfrac{\partial \mathbf{x}(t)}{\partial A_{i \: j}} + \overline{\mathbf{D}} \: \dfrac{\partial \overline{\mathbf{u}}(t)}{\partial A_{i \: j}} \\
		& = &  \mathbf{C} \: \dfrac{\partial \mathbf{x}(t)}{\partial A_{i \: j}} + \overline{\mathbf{D}} \: \dfrac{\partial \overline{\mathbf{u}}(t)}{\partial \mathbf{y}(t)} \dfrac{\partial \mathbf{y}(t)}{\partial A_{i \: j}} .
\end{array}
\label{Eq:JWLS_A1}
\end{equation}
The first term in the right-hand side of Eq.~(\ref{Eq:JWLS_A1}) is obtained by taking the derivative of the state relation in Eqs.~(\ref{Eq:GreyBoxStateSpace}) with respect to $A_{i \: j}$, that is
\begin{equation}
\begin{array}{r c l}
    \dfrac{\partial \mathbf{\dot{x}}(t)}{\partial A_{i \: j}} & = & \dfrac{\partial}{\partial A_{i \: j}} \left ( \mathbf{A} \: \mathbf{x}(t) + \overline{\mathbf{B}} \: \overline{\mathbf{u}}(t) \right) \\
     & = & \mathbf{A} \: \dfrac{\partial \mathbf{x}(t)}{\partial A_{i \: j}} + \mathbf{I}^{\: n_{s} \times n_{s}}_{i \: j} \: \mathbf{x}(t) + \overline{\mathbf{B}} \: \dfrac{\partial \overline{\mathbf{u}}(t)}{\partial A_{i \: j}} \\
		& = &  \mathbf{A} \: \dfrac{\partial \mathbf{x}(t)}{\partial A_{i \: j}} + \mathbf{I}^{\: n_{s} \times n_{s}}_{i \: j} \: \mathbf{x}(t) + \overline{\mathbf{B}} \: \dfrac{\partial \overline{\mathbf{u}}(t)}{\partial \mathbf{y}(t)} \dfrac{\partial \mathbf{y}(t)}{\partial A_{i \: j}} ,
\end{array}
\label{Eq:JWLS_A2}
\end{equation}
where $\mathbf{I}^{\: n_{s} \times n_{s}}_{i \: j}$ is a zero matrix with a single element equal to one at entry $\left( i,j \right)$.

The element $J_{A_{ij}}(t)$ is therefore given by the solution of the two equations
\begin{equation}
\left\lbrace
\begin{array}{r c l}
    \dfrac{\partial \mathbf{\dot{x}}(t)}{\partial A_{i \: j}} & = & \mathbf{A} \: \dfrac{\partial \mathbf{x}(t)}{\partial A_{i \: j}} + \mathbf{I}^{\: n_{s} \times n_{s}}_{i \: j} \: \mathbf{x}(t) + \overline{\mathbf{B}} \: \dfrac{\partial \overline{\mathbf{u}}(t)}{\partial \mathbf{y}(t)} \dfrac{\partial \mathbf{y}(t)}{\partial A_{i \: j}} \\
	  \dfrac{\partial \mathbf{y}(t)}{\partial A_{i \: j}} & = & \mathbf{C} \: \dfrac{\partial \mathbf{x}(t)}{\partial A_{i \: j}} + \overline{\mathbf{D}} \: \dfrac{\partial \overline{\mathbf{u}}(t)}{\partial \mathbf{y}(t)} \dfrac{\partial \mathbf{y}(t)}{\partial A_{i \: j}} .
\end{array} \right.
\label{Eq:JAij}
\end{equation}
Introducing the notations
$$ \begin{array}{c c}
\mathbf{x^{\ast}}(t) = \dfrac{\partial \mathbf{x}(t)}{\partial A_{i \: j}} \: ; & \mathbf{y^{\ast}}(t) = \dfrac{\partial \mathbf{y}(t)}{\partial A_{i \: j}} \: ; \\
\end{array} $$
\begin{equation}
\begin{array}{c}
 \overline{\mathbf{u}}^{\ast}(t) = \left( \begin{array}{c c}
             \mathbf{x}(t)^{T} & \left( \dfrac{\partial \overline{\mathbf{u}}(t)}{\partial \mathbf{y}(t)} \dfrac{\partial \mathbf{y}(t)}{\partial A_{i \: j}} \right)^{T} \\
             \end{array} \right)^{T} \\
\end{array}
\label{Eq:JAij_Notation1}
\end{equation}
and
$$ \begin{array}{c c}
\mathbf{A^{\ast}} = \mathbf{A} \: ; & \overline{\mathbf{B}}^{\ast} = \left( \begin{array}{c c}
             \mathbf{I}^{\: n_{s} \times n_{s}}_{i \: j} & \overline{\mathbf{B}} \\
             \end{array} \right);
\end{array} $$	
\begin{equation}
\begin{array}{c c}					
						\mathbf{C^{\ast}} = \mathbf{C} \: ; & \overline{\mathbf{D}}^{\ast} = \left( \begin{array}{c c}
             \mathbf{0}^{\: l \times n_{s}} & \overline{\mathbf{D}} \\
             \end{array} \right) ,
\end{array}
\label{Eq:JAij_Notation2}
\end{equation}
Eqs.~(\ref{Eq:JAij}) can be recast in the form
\begin{equation}
\left\lbrace
\begin{array}{r c l}
    \mathbf{\dot{x}^{\ast}}(t) & = & \mathbf{A^{\ast}} \: \mathbf{x^{\ast}}(t) + \overline{\mathbf{B}}^{\ast} \: \overline{\mathbf{u}}^{\ast}(t) \\
    \mathbf{y^{\ast}}(t) & = & \mathbf{C^{\ast}} \: \mathbf{x^{\ast}}(t) + \overline{\mathbf{D}}^{\ast} \: \overline{\mathbf{u}}^{\ast}(t) .
\end{array} \right.
\label{Eq:JAij_ast}
\end{equation}
Eqs.~(\ref{Eq:JAij_ast}) reveal that the elements of the Jacobian matrix associated with the parameters in $\mathbf{A}$ are solutions of an auxiliary state-space model defined by the four matrices $\left( \mathbf{A^{\ast}},\overline{\mathbf{B}}^{\ast},\mathbf{C^{\ast}},\overline{\mathbf{D}}^{\ast} \right)$. The first term in the auxiliary extended input $\overline{\mathbf{u}}^{\ast}(t)$ in Eq.~(\ref{Eq:JAij_Notation1}) is the state vector $\mathbf{x}(t)$. It is obtained by simulating in time the original model in Eqs.~(\ref{Eq:GreyBoxStateSpace}) with the estimated parameters of the previous Levenberg-Marquardt iteration. The second term in $\overline{\mathbf{u}}^{\ast}(t)$ depends on $\partial \overline{\mathbf{u}}(t) / \partial \mathbf{y}(t)$, which is formed using the derivatives of the nonlinear basis functions $\mathbf{g}(\mathbf{y}_{nl}(t),\mathbf{\dot{y}}_{nl}(t))$ with respect to $\mathbf{y}(t)$.

The determination of the element $J_{\overline{B}_{ij}}(t) \in \mathbb{R}^{\: l}$ is conducted similarly to $J_{A_{ij}}(t)$. The result is given in Eqs.~(\ref{Eq:JBij}), where $J_{\overline{B}_{ij}}(t)$ is seen to be the solution of another auxiliary state-space model,

\begin{equation}
\left\lbrace
\begin{array}{r c l}
    \dfrac{\partial \mathbf{\dot{x}}(t)}{\partial \overline{B}_{i \: j}} & = & \mathbf{A} \: \dfrac{\partial \mathbf{x}(t)}{\partial \overline{B}_{i \: j}} + \mathbf{I}^{\: n_{s} \times (m + s l)}_{i \: j} \: \overline{\mathbf{u}}(t) + \overline{\mathbf{B}} \: \dfrac{\partial \overline{\mathbf{u}}(t)}{\partial \mathbf{y}(t)} \dfrac{\partial \mathbf{y}(t)}{\partial \overline{B}_{i \: j}} \\
		& & \\
     \dfrac{\partial \mathbf{y}(t)}{\partial \overline{B}_{i \: j}} & = & \mathbf{C} \: \dfrac{\partial \mathbf{x}(t)}{\partial \overline{B}_{i \: j}} + \overline{\mathbf{D}} \: \dfrac{\partial\overline{\mathbf{u}}(t)}{\partial \mathbf{y}(t)} \dfrac{\partial \mathbf{y}(t)}{\partial \overline{B}_{i \: j}} .
\end{array} \right.
\label{Eq:JBij}
\end{equation}

The computation of $J_{C_{ij}}(t) \in \mathbb{R}^{\: l}$ and $J_{\overline{D}_{ij}}(t) \in \mathbb{R}^{\: l}$ is easier because they do not involve time integration, as shown in Eq.~(\ref{Eq:JCij}) and Eq.~(\ref{Eq:JDij}), respectively,

\begin{equation}
\begin{array}{r c l}
    \dfrac{\partial \mathbf{y}(t)}{\partial C_{i \: j}} & = & \mathbf{I}^{\: l \times n_{s}}_{i \: j} \: \mathbf{x}(t) + \overline{\mathbf{D}} \: \dfrac{\partial \overline{\mathbf{u}}(t)}{\partial \mathbf{y}(t)} \dfrac{\partial \mathbf{y}(t)}{\partial C_{i \: j}} ;
\end{array}
\label{Eq:JCij}
\end{equation}

\begin{equation}
\begin{array}{r c l}
    \dfrac{\partial \mathbf{y}(t)}{\partial \overline{D}_{i \: j}} & = & \mathbf{I}^{\: l \times (m + s l)}_{i \: j} \:\overline{\mathbf{u}}(t) + \overline{\mathbf{D}} \: \dfrac{\partial \overline{\mathbf{u}}(t)}{\partial \mathbf{y}(t)} \dfrac{\partial \mathbf{y}(t)}{\partial \overline{D}_{i \: j}} .
\end{array}
\label{Eq:JDij}
\end{equation}

\section{Experimental demonstration \mbox{on the Silverbox benchmark}}

\begin{figure}[ht]
\begin{center}
\includegraphics[width=75mm]{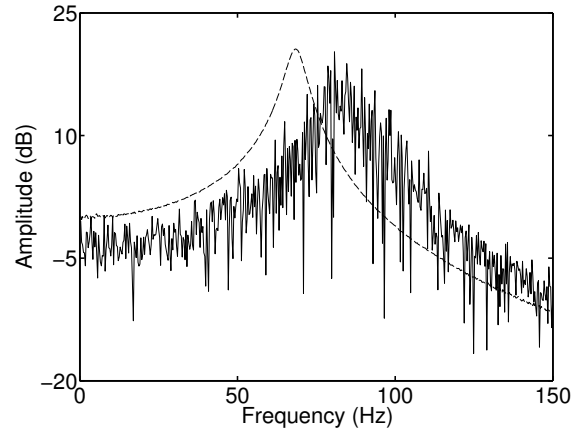} \\
\caption{Comparison of FRFs measured at 5 (dashed line) and 150 (solid line) $mN$ RMS.}
\label{Fig:Silverbox_FRF}
\end{center}
\end{figure}

The identification procedure described in Section 3 is demonstrated herein using experimental measurements acquired on the Silverbox circuit mimicking the behaviour of a single-degree-of-freedom nonlinear mechanical system. Ideally, this system should exhibit the dynamics of a Duffing oscillator with cubic nonlinearity, as prescribed by the equation
\begin{equation}
M \: \ddot{q}(t) + C_{v} \: \dot{q}(t) + K \: q(t) + c_{1} \: q^{3}(t) = p(t) .
\label{Eq:Duffing}
\end{equation}
In practice, it is also known to be characterised by, at least, an additional quadratic stiffness term $c_{2} \: q^{2}(t)$. The system was excited using random phase multisines~[\cite{JSchoukens_Book}] considering equivalent root-mean-squared (RMS) amplitudes of 5 and 150 $mN$. The input frequency spectrum was limited to 0 -- 300 $Hz$, excluding the DC component, with a sampling frequency of 2441 $Hz$. Experiments were conducted over 25 periods of 8192 samples, removing the first 5 periods to achieve steady-state conditions. Table~\ref{Table:Silverbox} reports the underlying linear modal properties of the benchmark estimated using a subspace analysis at 5 $mN$ RMS. Fig.~\ref{Fig:Silverbox_FRF} depicts the comparison between frequency response functions (FRFs) measured at 5 and 150 $mN$ RMS. The two curves reveal that the Silverbox vibrates in a strongly nonlinear regime of motion at high level, as a shift of the resonance frequency of more than 13 $Hz$ is noticed at 150 $mN$ RMS together with severe noisy-like distortions.

\begin{table}[ht]
\begin{center}
\begin{tabular}{c c}
\hline
Natural frequency ($Hz$) & Damping ratio ($\%$) \\
68.57 &  4.68  \\
\hline
\end{tabular}
\caption{Natural frequency and damping ratio of the Silverbox estimated at 5 $mN$ RMS.} 
\label{Table:Silverbox}
\end{center}
\end{table}

A state-space model in the grey-box form of Eqs.~(\ref{Eq:GreyBoxStateSpace}) is constructed using as nonlinear terms quadratic and cubic functions of the measured output displacement $y_{nl}(t)$. A model order equal to 2 is obviously selected and the resulting vector $\theta$ in Eq.~(\ref{Eq:theta}) thus consists of 15 parameters. Comparatively, in Ref.~[\cite{Paduart_PNLSS}], a black-box, second-order, state-space model, as in Eqs.~(\ref{Eq:BlackBoxStateSpace}), was adopted considering a third-degree multivariate polynomial in the state equation with all cross products included and linear terms only in the output equation. This led to a nonlinear model with 37 parameters.

\begin{figure}[ht]
\begin{center}
\includegraphics[width=75mm]{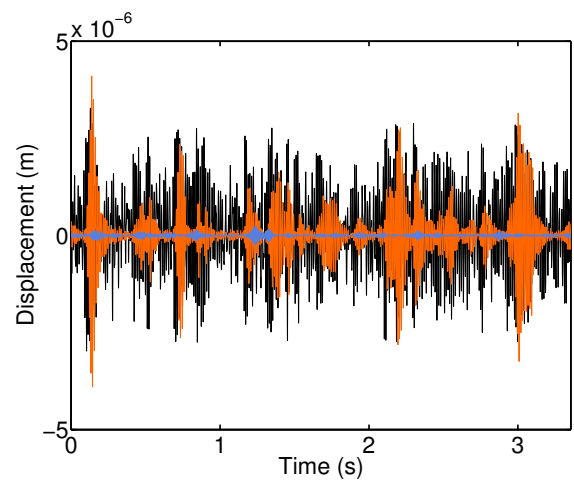} \\
\caption{Time-domain errors: measured response (black); initial model error (orange); final model error (blue).}
\label{Silverbox_TimeS}
\end{center}
\end{figure}

\begin{figure*}[p]
\begin{center}
\includegraphics[width=140mm]{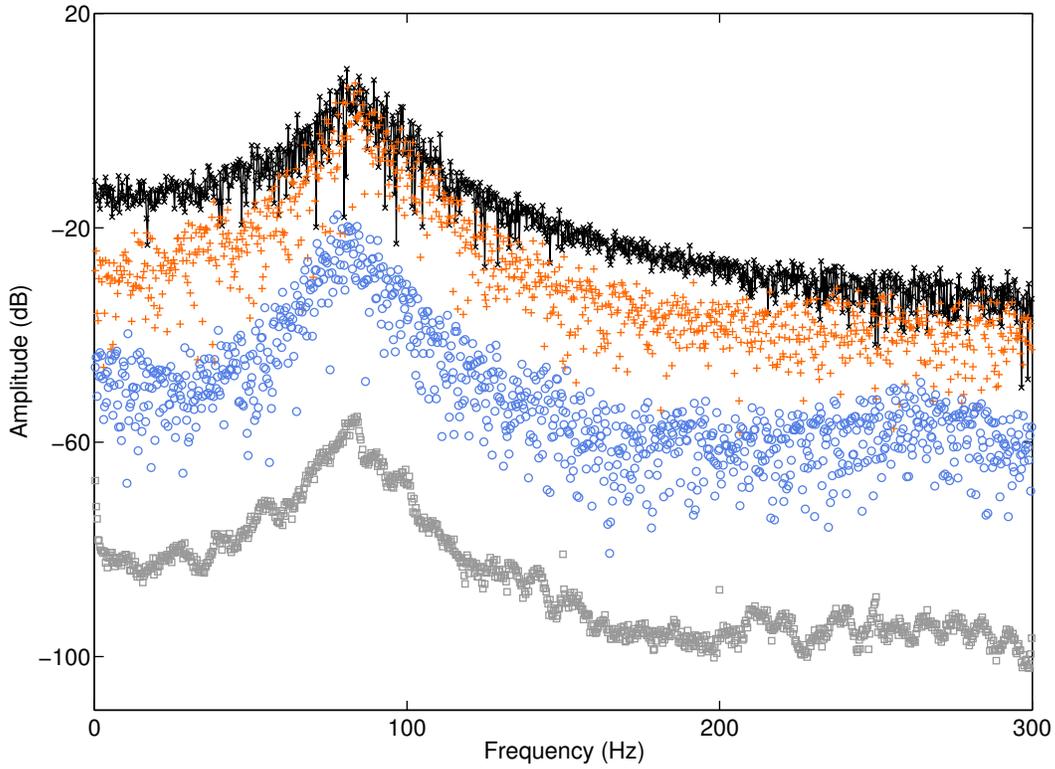} \\
\caption{Frequency-domain errors over 0 -- 300 $Hz$: measured spectrum (black cross); initial subspace model error (orange plus); final ML model error (blue circle); noise level (grey square).}
\label{Silverbox_OutputS}
\end{center}
\end{figure*}

The time- and frequency-domain errors associated with the two steps of the identification procedure are plotted in Figs.~\ref{Silverbox_TimeS} and~\ref{Silverbox_OutputS}, respectively. In the time domain in Fig.~\ref{Silverbox_TimeS}, the RMS error of the initial subspace model is equal to 6.24 $10^{-7}$ $m$, compared to the signal RMS value of 10.54 $10^{-7}$ $m$. The error is decreased down to 0.40 $10^{-7}$ $m$ after ML optimisation. The analysis of the reconstructed spectra in Fig.~\ref{Silverbox_OutputS} is also interesting. The error of the final model is generally 30 $dB$ below the measured output spectrum and 20 $dB$ lower than the subspace model. However, it does not reach the noise level, which is most probably due to an imperfect representation of the nonlinearity in the system. Errors in the final model are particularly visible in Fig.~\ref{Silverbox_OutputS} at the resonance location around 83 $Hz$, and close to third harmonics of the system around 250 $Hz$, proving that they are related to the modelling of the nonlinearity. 

\begin{figure*}[p]
\begin{center}
\begin{tabular}{c c}
\includegraphics[width=75mm]{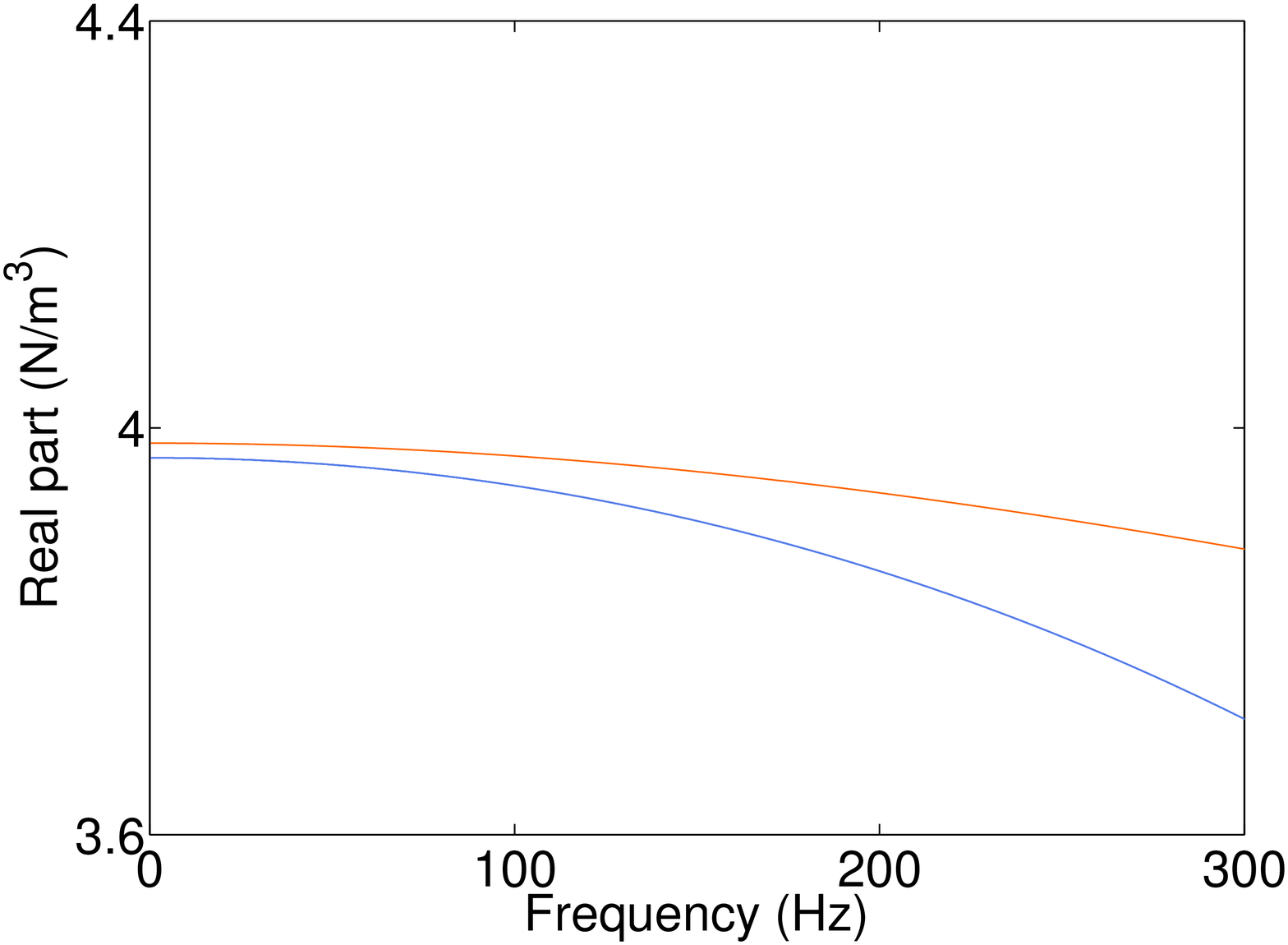} &
\includegraphics[width=75mm]{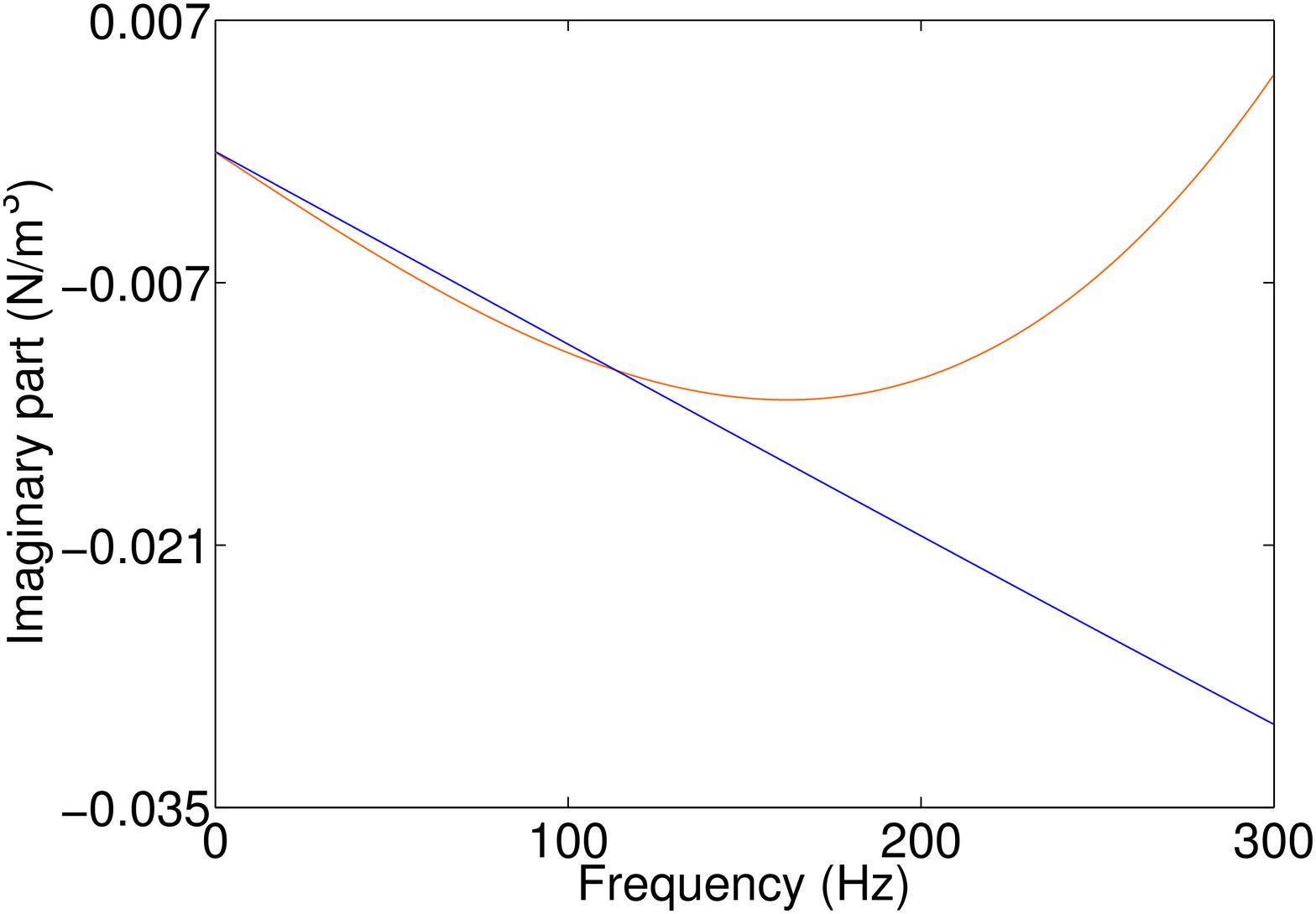} \\
\includegraphics[width=75mm]{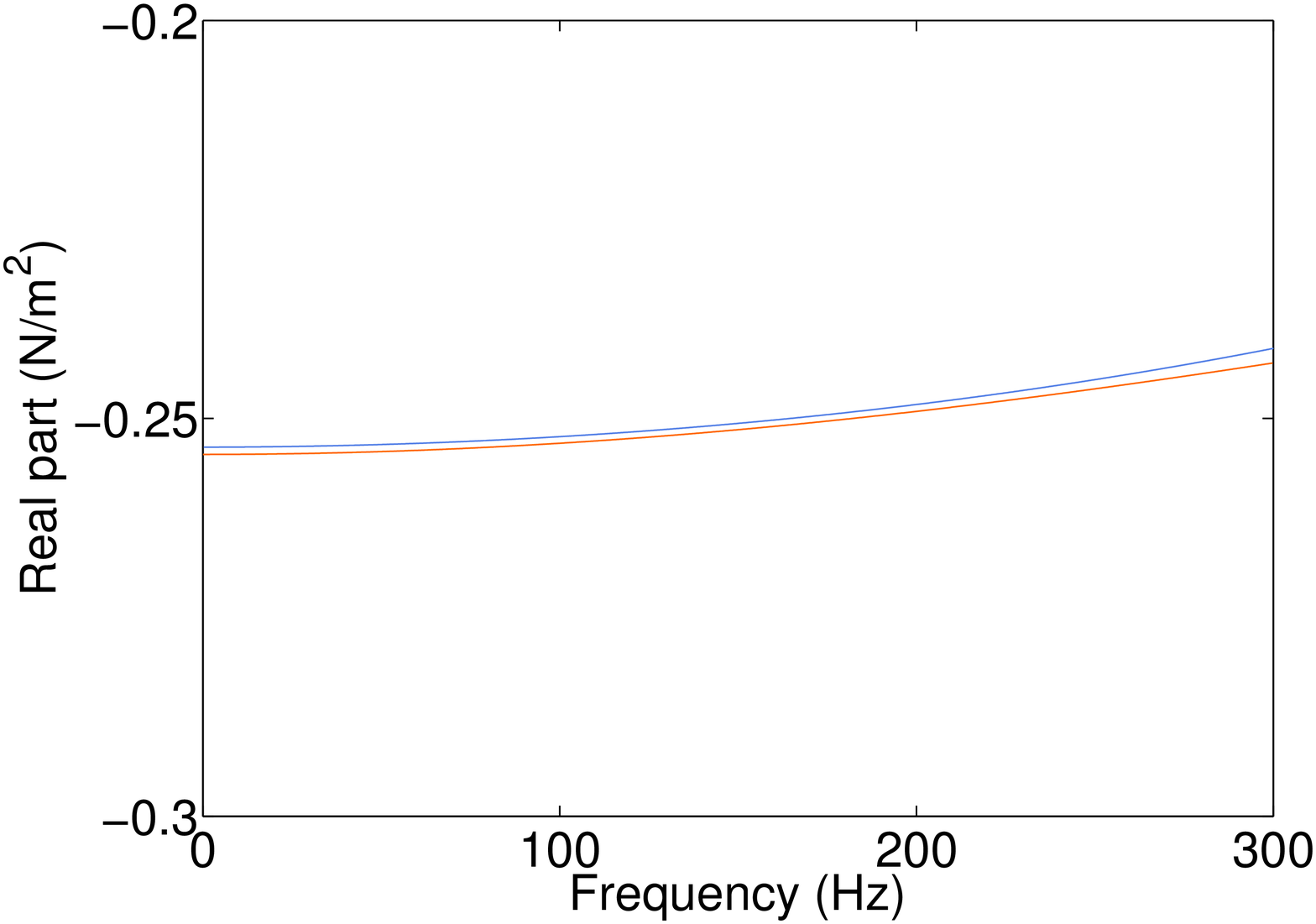} &
\includegraphics[width=75mm]{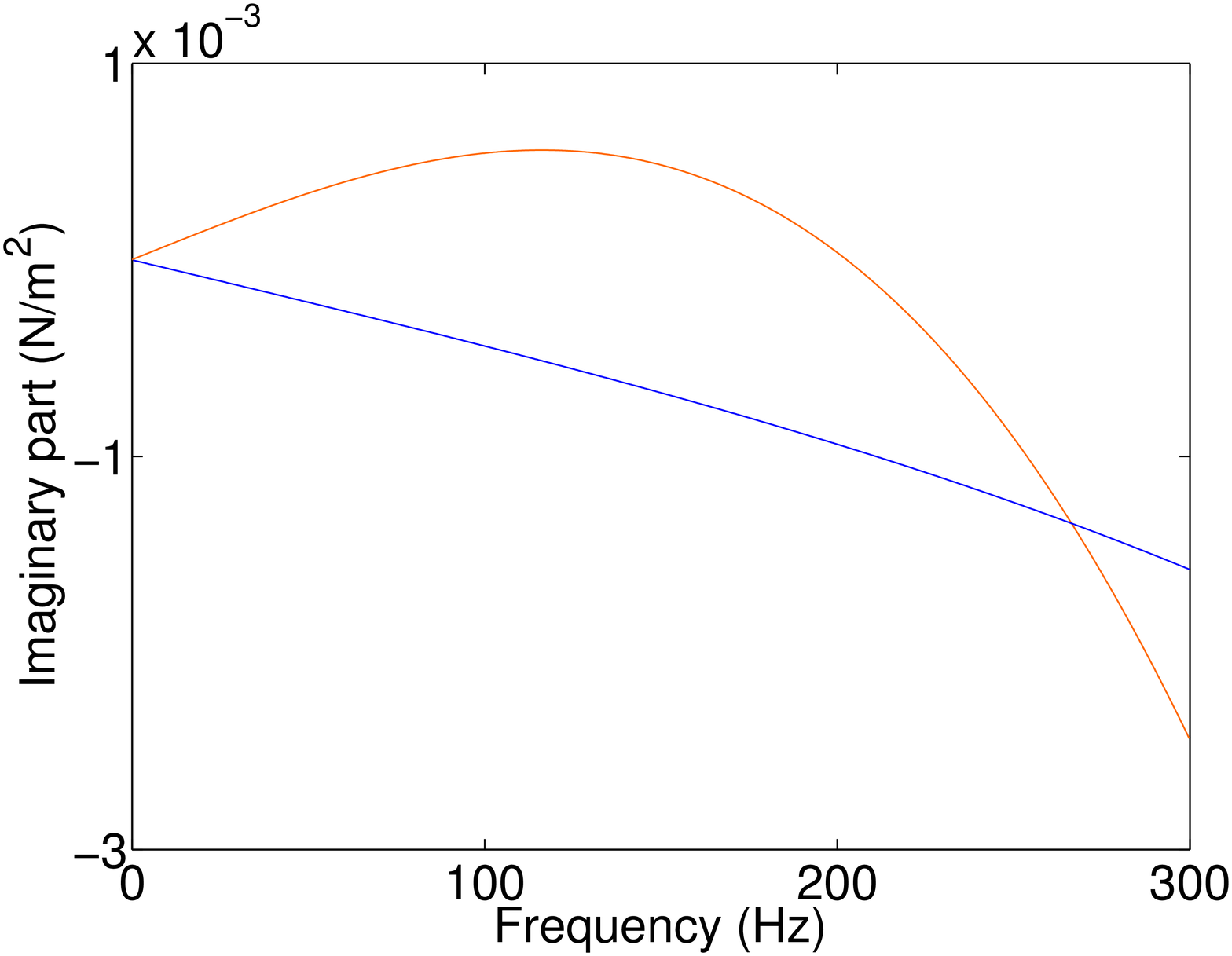} \\
\end{tabular}
\caption{Complex and frequency-dependent nonlinear coefficients $c_{1}$ (top left -- right) and $c_{2}$ (bottom left -- right) obtained by converting the initial (in orange) and final (in blue) state-space model parameters.}
\label{Fig:Silverbox_knl}
\end{center}
\end{figure*}

The final estimates of the state-space parameters can be converted into estimates of the coefficients $c_{1}$ and $c_{2}$ associated with the cubic and quadratic basis functions in the model, respectively. This is achieved using the conversion scheme proposed in Ref.~[\cite{Marchesiello_TNSI}], which yields frequency-dependent and complex-valued coefficients. As a result of the consistency of the identification procedure of Section 3, and in the absence of modelling errors, one expects the real parts of the coefficients to converge asymptotically to their exact values with no frequency dependence, and the imaginary parts to converge accordingly to zero. This makes the significance of the frequency variations and imaginary parts of the coefficients a particularly convenient means to assess the quality of the identification results.

The estimated nonlinear coefficients $c_{1}$ and $c_{2}$ are displayed versus frequency in Fig.~\ref{Fig:Silverbox_knl}~(a -- d). Table~\ref{Table:Silverbox_Results} lists the frequency averages of their real parts and the ratios between real and imaginary parts in logarithmic scaling. The real parts of the coefficients given by the initial and final models are almost equal. They are also found to be satisfactorily stable versus frequency, and remain more than 2 orders of magnitude greater than the corresponding imaginary parts. It should however be noted that the substantial decrease of the error from the initial to the final model in Figs.~\ref{Silverbox_TimeS} and~\ref{Silverbox_OutputS} does not translate into noticeably improved estimates of the nonlinear coefficients. This deserves more investigation to precisely understand the relation between the quality of physical-space parameters and the overall quality of the grey-box state-space model.

\begin{table}[ht]
\begin{center}
\begin{tabular}{c c c}
\hline
 & Initial subspace model & Final model \\
Real part $c_{1}$ ($N/m^{3}$) & 3.95 & 3.93 \\
$\mbox{Log}_{10}$ (real/imag.) & 2.71 & 2.41 \\
Real part $c_{2}$ ($N/m^{2}$) & -0.25 & -0.25 \\
$\mbox{Log}_{10}$ (real/imag.) & 3.38 & 2.54 \\
\hline
\end{tabular}
\caption{Estimates of the nonlinear coefficients $c_{1}$ and $c_{2}$ obtained by converting the initial and final state-space model parameters.} 
\label{Table:Silverbox_Results}
\end{center}
\end{table}

\section{Conclusion}

The objective of the present paper was to propose a grey-box state-space modelling framework to support the identification of nonlinear mechanical vibrations. It was shown that this framework paves the way for an important decrease of the number of model parameters with respect to classical black-box state-space modelling. The Silverbox benchmark was considered as an experimental case study demonstrating the derived identification procedure, which combines nonlinear subspace parameter initialisation and maximum likelihood optimisation. 

Additional work should focus on studying the sensitivity of the final parameter estimates to the amplitude of excitation and to the quality of the initial subspace model. The convergence of the parameters throughout the maximum likelihood iterations should also be analysed in more details. More advanced research prospects include the introduction of a spline-based representation of nonlinearities in the grey-box framework and the calculation of reliable confidence bounds on the model parameters.

\begin{ack}

The author J.P. No\"el is a Postdoctoral Researcher of the \textit{Fonds de la Recherche Scientifique -- FNRS} which is gratefully acknowledged. This work was also supported in part by the Fund for Scientific Research (FWO-Vlaanderen), by the Flemish Government (Methusalem), by the Belgian Government through the Inter university Poles of Attraction (IAP VII) Program, and by the ERC advanced grant SNLSID, under contract 320378.

\end{ack}

\bibliography{Bibliography_JPNoel}             
                                                   

\end{document}